\documentclass[a4paper]{article}
\usepackage[utf8x]{inputenc}
\usepackage[english]{babel}
\usepackage{amsmath,amsfonts,amssymb,amsthm}
\usepackage[sc]{mathpazo} 
\usepackage{microtype} 
\usepackage{graphicx}
\usepackage[hidelinks]{hyperref}
\usepackage[colorinlistoftodos,prependcaption,textsize=tiny]{todonotes}
\usepackage{rotating}
\usepackage{authblk}
\begin{document}
\date{}

\title{Community detection in weighted brain connectivity networks beyond the resolution limit}
\author[1,2]{\rm Carlo Nicolini}
\author[1]{\rm C\'ecile Bordier}
\author[1]{\rm Angelo Bifone}
\affil[1]{Center for Neuroscience and Cognitive Systems, Istituto Italiano di Tecnologia, Rovereto (TN), Italy}
\affil[2]{University of Verona, Verona, Italy}

\maketitle

\begin{abstract}
Graph theory provides a powerful framework to investigate brain functional connectivity networks and their modular organization. However, most graph-based methods suffer from a fundamental resolution limit that may have affected previous studies and prevented detection of modules, or “communities”,  that are smaller than a specific scale. Surprise, a resolution-limit-free function rooted in discrete probability theory, has been recently introduced and applied to brain networks, revealing a wide size-distribution of functional modules ~\cite{Nicolini2016}, in contrast with many previous reports. However, the use of Surprise is limited to binary networks, while brain networks are intrinsically weighted, reflecting a continuous distribution of connectivity strengths between different brain regions. Here, we propose Asymptotical Surprise, a continuous version of Surprise, for the study of weighted brain connectivity networks, and validate this approach in synthetic networks endowed with a ground-truth modular structure. We compare Asymptotical Surprise with leading community detection methods currently in use and show its superior sensitivity in the detection of small modules even in the presence of noise and intersubject variability such as those observed in fMRI data. Finally, we apply our novel approach to functional connectivity networks from resting state fMRI experimenta, and demonstrate a heterogeneous modular organization, with a wide distribution of clusters spanning multiple scales.
\end{abstract}

\section{Introduction}
The brain is thought to consist of a network of interconnected, interacting  components whose architecture is critical for the emergence of adaptive behaviors and cognition~\cite{mcintosh2000}. 
Graph theory provides a powerful means to assess topology and organization of brain connectivity networks, like those derived from MRI and other neuroimaging methods~\cite{eguiluz2005,bullmore2009}.
Within this framework, the brain is represented as a network of $n$ nodes interconnected by $m$ links.
Typically, the nodes correspond to anatomically defined brain regions and the links to a measure of interregional interaction or similarity~\cite{bullmore2009}. For resting state functional connectivity networks, edge weights are defined as interregional temporal correlations in the fluctuations of the BOLD signals, and the resulting graph can be represented by a correlation adjacency matrix.
The arcs of structural connectivity networks (the ``connectome''), conversely, reflect the number of white matter tracts connecting any two regions. Brain networks have also been defined on the basis of intersubject anatomical covariance~\cite{Evans2013}, co-activation of different brain regions across individuals subjected to experimental tasks~\cite{crossley2013a} or pharmacological challenges~\cite{Schwarz2007,schwarz2008}. All of these networks are ``weighted'' by definition, i.e. their edges are associated with real numbers representing a measure of the strength of pairwise interactions between nodes.

Graph–theoretical analysis of these networks has contributed substantially to our understanding of the topological organization of brain connectivity, revealing a small-world, rich-club structure~\cite{eguiluz2005,VandenHeuvel2011} and the presence of hub regions characterized by high connectivity and network centrality. Additionally, a number of studies (reviewed in~\cite{bullmore2009,vandenheuvel2010}) have investigated the modular structure of brain connectivity networks, highlighting cohesive clusters of nodes that are more densely connected among themselves than with the rest of the network. In the graph-theory jargon, these disjoint clusters are sometimes dubbed ``communities'', remnant of early investigations in the field of social sciences~\cite{Girvan2002a}.

Topological modularity is thought to reflect functional and anatomical segregation, a feature that may confer robustness and adaptivity to brain networks. 
Moreover, the degree of clustering within functional connectivity graphs may provide a measure of the balance between segregation and integration underlying brain function~\cite{bullmore2009}.
Finally, the identification of modules and their boundaries is important to understand the topological function of hub regions within the network~\cite{meunier2010}. 
Indeed, hubs sharing a large number of within-module edges may be critical to determine segregation of sub-structures within the network, while hubs connecting different modules are responsible for network integration~\cite{vandenHeuvel2013a}. Alterations in the community  structure of the brain have been observed in several neuropsychiatric conditions, including Alzheimer disease~\cite{Tijms2013}, schizophrenia~\cite{Stam2014} and chronic pain~\cite{Balenzuela2010}, and assessment of the brain modular organization may provide a key to understanding the relation between aberrant connectivity and  brain disease.

Following initial work by~\cite{hilgetag2000a}, several graph theoretical methods have been deployed to investigate the modular structure of brain networks~\cite{meunier2009,meunier2010,Power2011}.
Typically, these methods rely on the optimization of a fitness function that measures the quality of a network partition against that of an ensemble of randomized networks with similar statistical properties (the ``null model''). Optimization of the fitness function of choice is often computationally demanding and scales steeply with increasing network size. Hence, heuristics are needed to calculate nearly optimal partitions of large networks, like those derived from neuroimaging data, within reasonable computation time~\cite{blondel2008,rosvall2008}.

A seminal finding in graph theory is that clustering methods based on optimization of a global function suffer from a resolution limit~\cite{fortunato2007}, as they are unable to resolve modules that are smaller than a scale determined by the size of the entire network. This problem was first demonstrated for Newman's Modularity~\cite{newman2006}, a method included in the Brain Connectivity Toolbox~\cite{rubinov2010} and most frequently applied to the analysis of neuroimaging data.

Subsequent work by various groups has demonstrated that the resolution limit is quite pervasive and affects, to a different extent, many other methods based on optimization of a global fitness functions~\cite{squartini2015,traag2011,lancichinetti2009}, including Reichardt and Bornholdt's~\cite{reichardt2006}, Arenas and Gomez'~\cite{arenas2008}, Ronhovde and Nussinov's~\cite{ronhovde2009}, Rosvall and Bergstrom's (\emph{Infomap})~\cite{rosvall2008,kawamoto2015} and others.

The introduction of a resolution parameter has been proposed as a means to mitigate the problem by adjusting the resolving power of the function to a specific scale~\cite{reichardt2006,ronhovde2010,yeo2011}. However, this approach enables resolution of smaller clusters at the expense of larger ones, which may be unduly subdivided, thus resulting in partitions with relatively uniform cluster size distributions that do not capture the complex modular structure of real-world networks~\cite{lancichinetti2011}. 

Recently, we have assessed the effects of the resolution limit on the analysis of brain connectivity networks~\cite{Nicolini2016}. Specifically, we have shown that this limitation severely curtails the ability to detect small, but functionally and anatomically meaningful clusters of nodes even when they present high densities of intra-cluster edges. Moreover, we showed that resolution-limited methods, like Newman's Modularity, do not reflect the multiple scales of the organization of brain connectivity networks, where small and large modules can coexist. 
We have also demonstrated that Surprise, a conceptually different fitness function grounded in probability theory, behaves like a resolution-limit-free function~\cite{Nicolini2016}. Maximization of Surprise, based on an algorithm dubbed FAGSO, revealed a heterogeneous distribution of modules within brain resting state and coactivation  networks. If confirmed, these findings would suggest that a substantial revision of current models of brain modular structure may be in order.

A fundamental limitation of Surprise lies in its definition in terms of discrete probability and binomial coefficients that make it applicable only to binary networks, i.e. graphs with edge values $1$ or $0$. This may represent a substantial drawback, for it requires binarization of brain connectivity networks, thus discarding potentially important information contained in the edge weight distribution.
Moreover, different binarization procedures may lead to different network representations for the same connectivity dataset. Therefore, an extension of Surprise to weighted networks would be highly desirable, and would provide a new and important tool to study the modular organization of brain connectivity beyond the resolution limit.

Capitalizing on recent development in the field of statistical physics of complex networks~\cite{traag2015}, here we describe and demonstrate the use of Asymptotical Surprise, a weighted counterpart to Surprise, in the study of the modular structure of weighted networks.

Moreover, we propose a new algorithm, dubbed PACO (PArtitioning Cost Optimization) for the maximization of Asymptotical Surprise.
The performance of this novel approach is assessed on synthetic networks with pre-defined ground-truth modular structures, and compared to some of the leading graph partitioning methods.
Importantly, we demonstrate our approach in networks  derived from synthetic data that mimic different structures, levels of noise and variability, such as those observed in functional connectivity experimental data.
Indeed, improved resolution afforded by Asymptotical Surprise may imply  increased vulnerability to spurious modules resulting from noisy correlations.
It is therefore important to assess the benefits of increased resolution against the limitations arising from intrinsic data variability. 

Finally, we apply Asymptotical Surprise to weighted functional connectivity networks from resting state fMRI data. The implications of the heterogeneous, multiscale community structure revealed by this resolution-limit-free method are discussed in the context of the current models of the brain modular organization.

\section{Materials and Methods}
\subsection{Notation}
Here we briefly summarize the terminology and the notation that will be used throughout the paper.
A binary graph $G=(V,E)$ is a representation of a set $V$ of $n$ nodes, also called \emph{vertices}, connected by $m$ links (or edges), in a set $E$. The adjacency matrix $A=\{a_{ij}\}$ of a binary graph is a square $n\times n$ symmetric matrix with elements $A_{ij}=1$ when an edge exists between vertex $i$ and $j$ and $0$ otherwise. We denote the total number of possible links in the graph as $p=\binom{n}{2}$.

A weighted graph $G=(V,E,W)$ assigns as a set of edge weights $W$ to the links. For weighted graphs, the adjacency matrix is square, symmetrical and has real elements.

A clustering $\zeta = \{\zeta_c\}$ of $G$ is a partitioning of $V$ into disjoint sets of nodes, $\zeta_c \subseteq V$, which we call modules or communities. Each module consists of $n_c$ nodes, $m_c$ edges and $p_c=\binom{n_c}{2}$ pairs of nodes. On weighted graphs we define $m_c$ as the sum of edge weights inside a community.
The sum of edges internal to all communities, $m_\zeta$ and the intra-cluster pairs $p_\zeta$ are $m_\zeta = \sum_c m_c$ and $p_\zeta=\sum_c p_c$ respectively.


\subsection{Surprise and Asymptotical Surprise}
Surprise~\cite{aldecoa2011,aldecoa2013} is a quality measure of the partition of a binary network that has its roots in probability theory.
For a given partition $\zeta$, Surprise represents the probability that a graph drawn uniformly at random from the set of all graphs with $n$ nodes, $p=\binom{n}{2}$ pairs and $m$ edges has at least as many intra-cluster edges as $G$. Intuitively the lower the probability the better the partition.

For binary networks, Surprise can be computed within the discrete probability theory of urn models as: 
\begin{equation}
S = \sum_{i = m_\zeta}^m \dfrac{\binom{p_\zeta}{i} \binom{p-p_\zeta}{m-i} }{\binom{p}{m}}
\end{equation}
Due to numerical precision problems in the evaluation of large binomial coefficients, $\hat{S}(\zeta)=-\log_{10} S(\zeta)$ is often taken as measure of quality of the partition, with higher values corresponding to better clustering.

Surprise quantifies the extent of the departure of the distribution of intra-cluster nodes and edges from that of a randomly drawn partition with the same internal density as in the original graph~\cite{alba1973}.
In the limit of large networks, Surprise can be approximated by a binomial distribution: this observation led to weighted definition of Surprise $\hat{S}$, dubbed Asymptotical Surprise~\cite{traag2015}:

\begin{equation}\label{eq:asymptoticalsurprise}
\mathcal{S}_a = m D_{\textrm{KL}}\left( q \| \left< q \right> \right)
\end{equation}

where, for brevity of notation, $q=m_\zeta/m$ and $\left< q \right> = p_\zeta/p$ are the observed and expected fraction of intra-cluster links relatively and $D_{\textrm{KL}}(x\|| y) = x \log(x/y) + (1-x)\log((1-x)/(1-y))$ is the binary Kullback-Leibler divergence~\cite{kullback1951}.

In the framework of information theory~\cite{cover2006}, Asymptotical Surprise represents the Kullback-Leibler divergence between the observed and expected fraction of intra-cluster edges; it encodes the information lost when the prior distribution $\left <q \right >$ is used to approximate the posterior distribution $q$. Kullback-Leibler divergence is a quasi-distance on probability distributions as it is always non-negative, non-symmetric and zero only when $q=\left< q \right>$ like binary Surprise.

Asymptotical Surprise has a simpler formulation than binary Surprise as there are no binomial coefficients to evaluate and it has been shown to be resolution-limit-free in the limit of large networks ~\cite{traag2015}.

\subsection{Maximization of Asymptotical Surprise}
Finding the optimal partition of a graph is an NP-hard problem~\cite{fortunato2010} and practical implementations of community detection rely on heuristic approaches that enable finding nearly-optimal solutions in a reasonable computation time.

Here we introduce a powerful and general method for the optimization of Asymptotical Surprise dubbed PACO (PArtitioning Cost Optimization). PACO is a non-deterministic agglomerative algorithm based on FAGSO and, like the Louvain method, has an element of randomness that enables a more efficient exploration of the partition landscape.

The operating principle of PACO is based on the triadic closure property, i.e. the fact that in real-world networks nodes with many common neighbors are more likely to be neighbors. This transitive neighborhood property underlies the formation of communities of nodes~\cite{Eustace2015}. In principle, any measure of structural similarity between nodes could guide a community detection heuristic toward the optimal partition. Specifically, PACO uses the Jaccard index~\cite{jaccard1901}, a measure of the fraction of overlap between the neighbors in common between nodes, as the guiding principle for the agglomeration of similar nodes in the same community.

In the first phase of PACO, the Jaccard metric is evaluated for every edge. More formally, for an edge $e=(u,v)$ the Jaccard index is computed as $J(e)=\frac{|\Gamma(u) \cap \Gamma(v)|}{|\Gamma(u) \cup \Gamma(v)|}$ where $\Gamma(u)$ and $\Gamma(v)$ are the neighboring nodes of $u$ and $v$ respectively.

The agglomerative process starts with an initial partition where every vertex represents a community on its own. This partition has $n$ communities and no intra-cluster edges.
The edges of the graph are then ranked in decreasing order by their Jaccard index and iteratively, for every edge in the sorted list, endpoint nodes are merged only if they belong to different communities. In this case one of the two endpoints, selected by chance, is assigned to the other's endpoint community and the increment of Surprise is computed: if it is positive, the partition is updated together with the new value of Surprise (or Asymptotical Surprise), otherwise the algorithm proceeds to the next edge.  Detailed pseudocode for PACO is shown in the Supplementary Materials, Figure S1.

The main difference between PACO and its predecessor FAGSO is the data structure used to store the community structure. FAGSO maintains the community structure in a disjoint-set data structure and when one vertex is moved into another's community, the two modules are merged into one (Supplementary Materials, Figure S2).
Conversely, PACO moves single nodes between different communities, and never merges modules (Supplementary Materials, Figure S2, boxes C, D). This results in a more finely-grained optimization that allows a better exploration of the quality function landscape.

\subsection{Synthetic benchmark networks}
Here we introduce a theoretically sound method for the generation of synthetic FC networks that mimic properties of resting state fMRI networks, including noise and intersubject variability, while presenting a pre-determined ground-truth modular structure against which the performance of community detection algorithms can be tested.

The general idea is that, starting from an adjacency matrix with a given modular structure, we can generate time-courses for each of the nodes whose pairwise correlations reproduce the edge structure of the original matrix. Noise can be added to the time-courses, and the resulting correlation matrix will provide a noisy representation of the original one. This procedure can be repeated multiple times to produce different datasets that represent different "subjects" in the study.

\begin{sidewaysfigure}
\centering
\includegraphics[height=0.4\textheight]{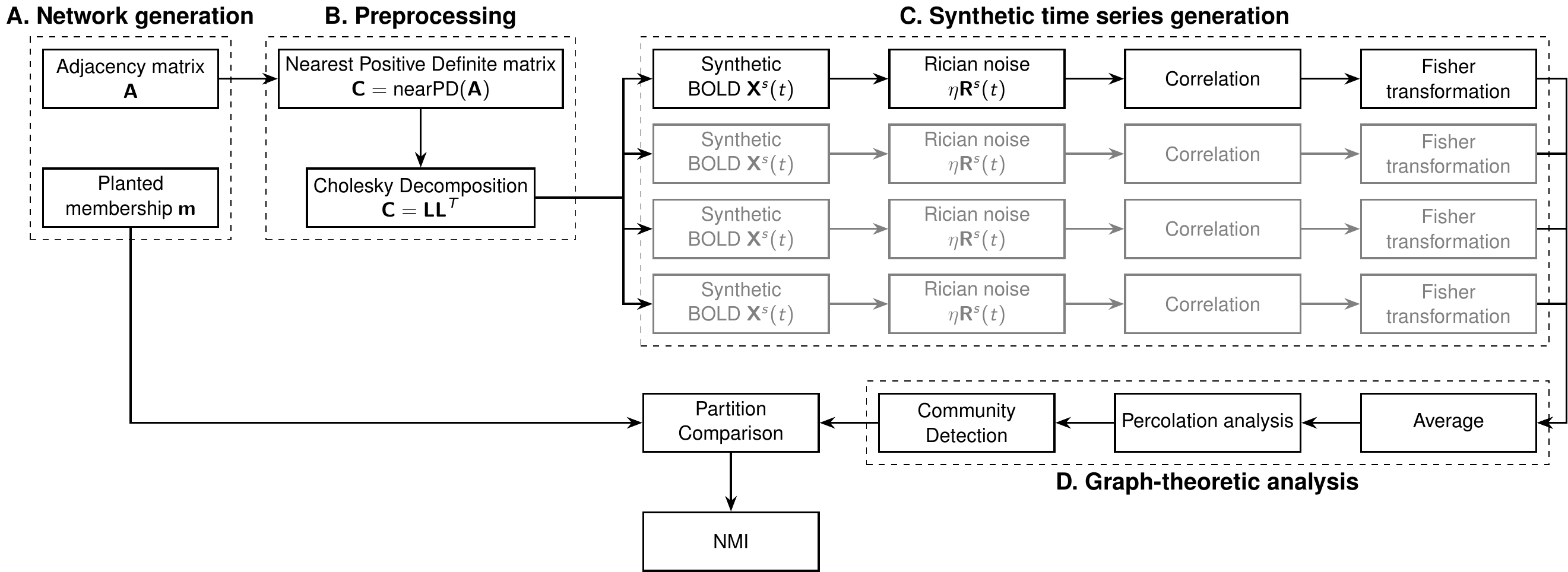}
\caption{Flowchart of the generation and analysis of the synthetic datasets. In A the network with a pre-defined community structure is generated. The adjacency matrix is then processed in block B to obtain the nearest positive definite matrix for the Cholesky decomposition. This enables the generation of node-wise time-courses into which different levels of noise can be injected. The procedure is repeated multiple times to generate different instances (mimicking different subjects in the sample). Finally, correlation matrices are calculated for each instance (block C), and Fisher transformed to calculate the average adjacency matrix for analysis by community detection algorithms (block D). Lastly, resulting partitions are compared with the original, planted one in terms of NMI.}
\label{fig:flowchart}
\end{sidewaysfigure}

In practical terms, given an undirected weighted graph  $\mathbf{C} \in \mathbb{R}^{n\times n}$ whose community structure is known a-priori, we have calculated its nearest positive definite matrix~\cite{Higham1988} and its Cholesky decomposition, i.e. an upper triangular matrix $\mathbf{L}\in \mathbb{R}^{n\times n}$ such that $\mathbf{L}\mathbf{L}^T=\mathbf{C}$. Starting from uncorrelated variables $\mathbf{X} \in \mathbb{R}^{n \times l}$, we have generated correlated random variables $\mathbf{Y}=\mathbf{L} \mathbf{X}$ such that $\mathbb{E}(\mathbf{Y}\mathbf{Y}^T)=\mathbf{C}$. Additionally, we have injected different levels of noise into $\mathbf{Y}$ prior to the computation of the correlation matrix. 
Schematic of this procedure is shown in Figure~\ref{fig:flowchart}.

We tested this idea on two different models of planted partition: a variant of the ring of cliques~\cite{fortunato2007} and the Lancichinetti-Fortunato-Radicchi (LFR) network~\cite{lancichinetti2008}, whose degree distribution and modular structure can be tuned to replicate topological features of real-world networks, including scale freeness~\cite{hagmann2008} and the presence of densely interconnected cores~\cite{VandenHeuvel2011}.

One important finding in~\cite{Nicolini2016} is that brain networks are organized in modules with heterogeneous size distributions.
We implemented this property in our two types of benchmark networks.
For the first test, we generated a ring of cliques with $300$ nodes, and sizes of the cliques sampled from a power-law with exponent $\tau_c=2$ minimum and maximum clique size respectively $\min_c=5$, $\max_c=50$.
For each subject of the sample, we synthesized $150$ time-points for each node using the \texttt{neuRosim R} package~\cite{neurosim2011}. We set the baseline value of all the time series to $100$~\cite{Welvaert2013}.

Finally, we correlated the original synthetic time series $\mathbf{X}$ by multiplication with the matrix $\mathbf{L}$, obtained the correlated time series $\mathbf{Y}$ and added Rician noise to $\mathbf{Y}$ independently for each area. The simulated data $\mathbf{Y}$ did not include slow drift components, simulated physiological noise, nor spatial noise. The average SNR was defined as $\textsc{SNR}=\bar{S}/\sigma_N$ where $\bar{S}$ is the average magnitude of the signal and $\sigma_N$ is the standard deviation of the noise~\cite{Kruger2011}.

In order to be more exhaustive and extend the validity of results, we repeated the same procedure on weighted LFR networks with $N=600$ nodes, sampling nodes degree from a power-law with exponent $\tau_d=2$, average degree $\left<k\right>=12$ and maximum degree $\max_k=50$.
We set the topological mixing coefficient, i.e. the fraction of intra-cluster and inter-cluster links, to $\mu_t=0.1$. Planted community sizes ranged from $5$ to $50$ nodes and were sampled from a power law with exponent $\tau_c=1$.


Group-level correlation matrices were computed by Fisher-transforming and averaging individual instances of the above matrices. Sparsification was obtained by removing edges with weights below a threshold determined by percolation analysis~\cite{Gallos2012,Bardella2016a}. This approach measures the size of the largest connected component of the network upon iterative removal of the weakest edges and enables data-driven determination of the optimal sparsification threshold that preserves network structure and connectedness while removing potentially spurious correlations.

\subsection{Comparative community detection methods}
The community structure of the resulting weighted sparsified matrices was detected by Asymptotical Surprise optimized with PACO and compared against two widely used methods, Infomap~\cite{rosvall2008} and Newman's Modularity~\cite{blondel2008,newman2006}, that  are affected by the resolution limit, albeit to different extents. 
In Newman's Modularity, the size of the smallest detectable cluster is of the order of the square root of the number of edges in the entire network~\cite{fortunato2007}. Infomap has a limit that depends on the overall number of inter-cluster edges~\cite{kawamoto2015}.

These two methods are based on different principles to detect the community structure of a graph.
Newman's Modularity finds the optimal partition by maximizing intra-cluster edge-density against that of a configuration model~\cite{newman2006}. Optimization of this fitness function is typically performed using the Louvain method, a greedy agglomerative clustering algorithm that works on hierarchical refinements of the network's partitions. Here we used the Louvain implementation available in the Brain Connectivity toolbox~\cite{rubinov2010}. 
The idea behind Infomap is the minimization of the description length~\cite{Rissanen1978} of a random walker defined on the network through a set of heuristics. For this study we used the Infomap implementation available in the \texttt{igraph-0.7.1} package~\cite{igraph2006}.

For all methods, including PACO, we launched $10,000$ independent runs, and picked the membership corresponding to the partition with the best value of the fitness function, the maximum for Modularity and Asymptotical Surprise, the minimum for Infomap.

Our implementation of PACO as well as the code to generate benchmark LFR networks was written in \texttt{C++} with bindings in Matlab, Octave, Python. The software is available upon request.

\subsection{Measures of partition quality}
For each method, we analyzed the level of agreement of the detected community structure against the planted one in terms of Normalized Mutual Information (NMI)~\cite{Danon2005}.
Additionally, we used two different coefficients of similarity between partitions: Sensitivity and Specificity. 

To this end, we quantified the confusion matrix $\mathbf{C}$ between the detected and planted modules. Each element $C_{ij}$ is the number of nodes in the planted community-$i$ that appear in the detected community-$j$.
For each planted community we scored as true positives (TP) the nodes correctly identified as belonging to the ground-truth community, and as false positives (FP) the nodes wrongly assigned to a community; similarly false negatives (FN) were nodes wrongly classified in different communities and true negatives (TN) the nodes correctly classified as out of the community.
Sensitivity, defined as $TP/(TP+FN)$, decreases with increasing number of False Negatives. Specificity instead is defined as $TN/(TN+FP)$ and decreases when many nodes are wrongly assigned in the same community.
Additionally, we computed Accuracy and Matthew Correlation Coefficient (see Supplementary Materials for definitions).

\subsection{Human resting state network}
We applied Asymptotical Surprise maximization by PACO to a reference resting state fMRI functional connectivity dataset from healthy subjects~\cite{crossley2013a} made available to the scientific community through the public Brain Connectivity Toolbox~\cite{rubinov2010}. 
Detailed experimental and image processing procedures are described in the original paper~\cite{crossley2013a}, alongside with the ethical statements.

In short, fMRI data were acquired from 27 healthy volunteers at 3T.
Gradient echo-planar imaging data were collected for 5 min with 2s TR and 13 and 31 ms echo-times. Thirty six interleaved 3mm slices with in-plane resolution of $3.5\times 3.5$ mm were acquired.
Time series were extracted from 638 brain regions defined by a template~\cite{crossley2013a}, corrected for motion and band-passed (0.01–0.1Hz). Functional connectivity was defined in terms of pairwise Pearson correlations at a subject's level.
A group-level functional connectivity matrix was calculated by averaging individuals' matrices after Fisher-transform, and thresholded to retain 18625 edges, as described in Crossley et al.~\cite{crossley2013a}.
We used BrainNetViewer as a tool for the visualization of the communities on brain templates~\cite{xia2013}.

\section{Results}
\subsection{Synthetic networks}
We compared the quality of the partitions of the synthetic benchmark networks  obtained by Asymptotical Surprise with those of Infomap~\cite{rosvall2008} and Newman's Modularity ~\cite{newman2006,blondel2008}. Figure \ref{fig:nmisensitivityspecificityringclique} shows Normalized Mutual Information, Sensitivity and Specificity of the three methods applied to the ring of cliques for different sample sizes and SNRs; no-noise condition is represented as ``Inf''.
This model network was constructed to test the ability of the three methods to retrieve heterogeneous community structures under various noise conditions.

As expected, all methods showed better performance with increasing SNR and number of subjects, as noise and intersubject variability introduce spurious edges that hinder the ability to retrieve the planted structure.
Partitions obtained with Newman's modularity showed the lowest NMI with respect to the planted partition under all conditions.
Sensitivity of Newman's modularity did not exceed $0.75$ even for high SNRs and a large number of subjects, a consequence of its stronger resolution limit.
For this network, Infomap performed substantially better in terms of NMI against the planted partition, with a Sensitivity that was superior to that of Modularity across the spectrum of conditions.

Asymptotical Surprise showed highest NMI and Sensitivity across conditions, consistent with its resolution-limit-free behavior. Asymptotical Surprise proved superior in terms of NMI and Sensitivity in the low SNR regimes, and in the presence of relatively large intersubject variability as mimicked by the generation of different instances of the ring of cliques (see Methods section). Specificity of Asymptotical Surprise was not inferior to the other methods under all conditions, thus ruling out increased vulnerability to False Positives, at least in this particular model network.

\begin{figure}[h!]
\includegraphics[width=1\textwidth]{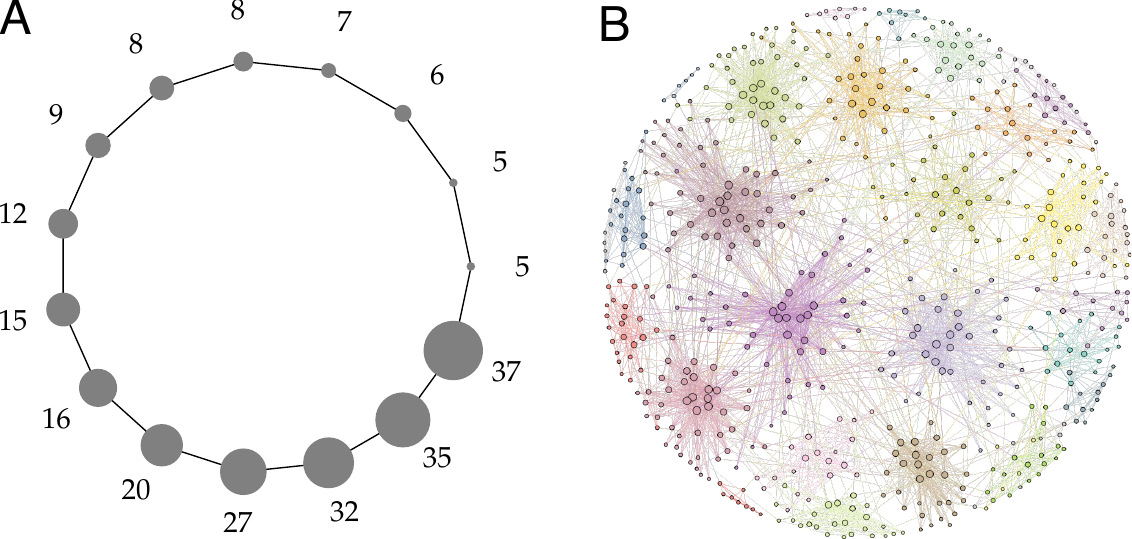}
\caption{The two benchmark networks used in this study, laid out. (A) is a power-law ring of cliques, where cliques present different sizes sampled from a power-law distribution;
(B) is the layout of an LFR network with parameters $N=600$, $\left< k \right>=12$,  $\max_k=50$, $\mu_t=0.1$, $\mu_w=0.1$, $\min_c=5$, $\max_c=50$.
The layout of (B) was generated with the graph-tool library~\cite{peixoto_graph_tool_2014}.}
\label{fig:lfrringclique}
\end{figure}

\begin{figure}[h!]
\centering
\includegraphics[width=\textwidth]{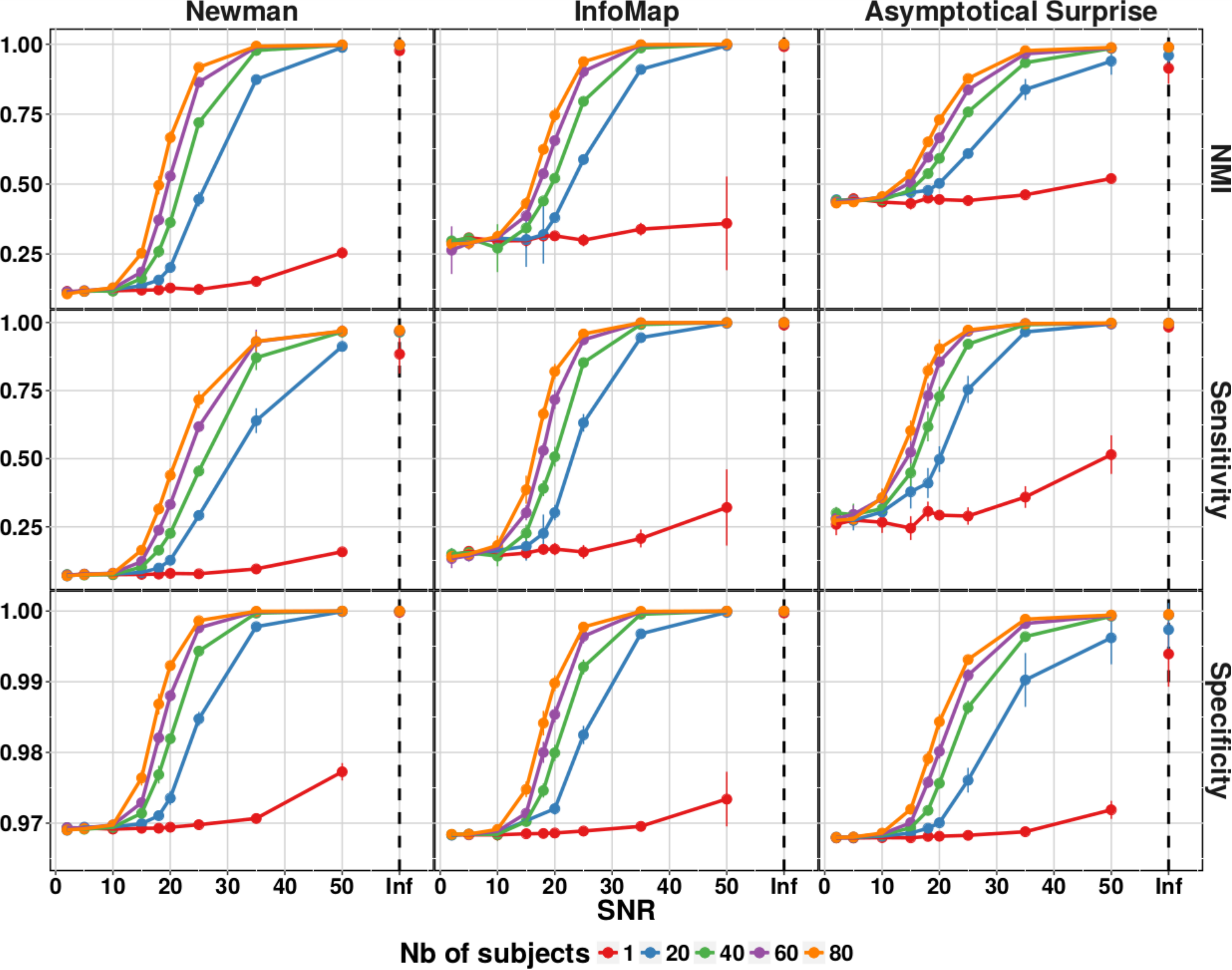}
\caption{NMI, Sensitivity and Specificity of the three community detection algorithms applied to Lancichinetti-Fortunato-Radicchi (LFR) networks. SNR indicates Signal to Noise Ratio, and Inf the situation with a network structure unperturbed by noise. Number of Subjects indicates the different number of instances used to generate the group level network.}
\label{fig:nmisensitivityspecificitylfr}
\end{figure}

Comparable results were obtained for the LFR network
(Figure~\ref{fig:nmisensitivityspecificitylfr}), a model graph that replicates the distribution of nodal degree observed in many real-world networks, including those representing brain functional connectivity.
All three methods showed similar values of NMI for high SNR and a large number of subjects, with a plateau reaching maximum Sensitivity with a group sample bigger than $20$ and SNR above $30$.
Sensitivity was only slightly worse for Modularity, but it should be noted that for the LFR network the size distribution of the planted modules was narrower than for the ring of cliques (Figure~\ref{fig:lfrringclique}), thus making the resolution limit less evident.

In the lower SNR regime, Asymptotical Surprise presented the best performance in terms of NMI and Sensitivity, with a slower decay for decreasing SNR. 
Specificity was almost equivalent across the three methods, with a quick convergence to the maximum value of 1 for high SNR and good performance (around $0.97$) for low SNR.
Asymptotical Surprise presented a faster decay with decreasing SNR.
However, it should be noticed that the scale of Specificity has a very narrow range ($0.97$-$1.00$), and the differences between the three methods were relatively small.

For the sake of completeness, we also computed Accuracy and Matthew Correlation Coefficient for the same model networks, shown in the Supplementary Materials.
Notably, Infomap showed a large variability in Accuracy for lower SNRs and number of subjects.
Under closer examination, however, it appeared that the increased variance for Infomap was due to occasional runs in which the algorithm only retrieved one or two large modules.
This is a known problem with Infomap and other algorithms based on random walks that depends on the need to parametrize the teleportation step in order to make the dynamics ergodic~\cite{lambiotte2012}.

Altogether, the picture that emerges from the analysis of Accuracy and MCC is entirely consistent with the results shown in this section.

\subsection{Resting state functional connectivity dataset}

\begin{figure}[h!]
\centering
\includegraphics[width=\textwidth]{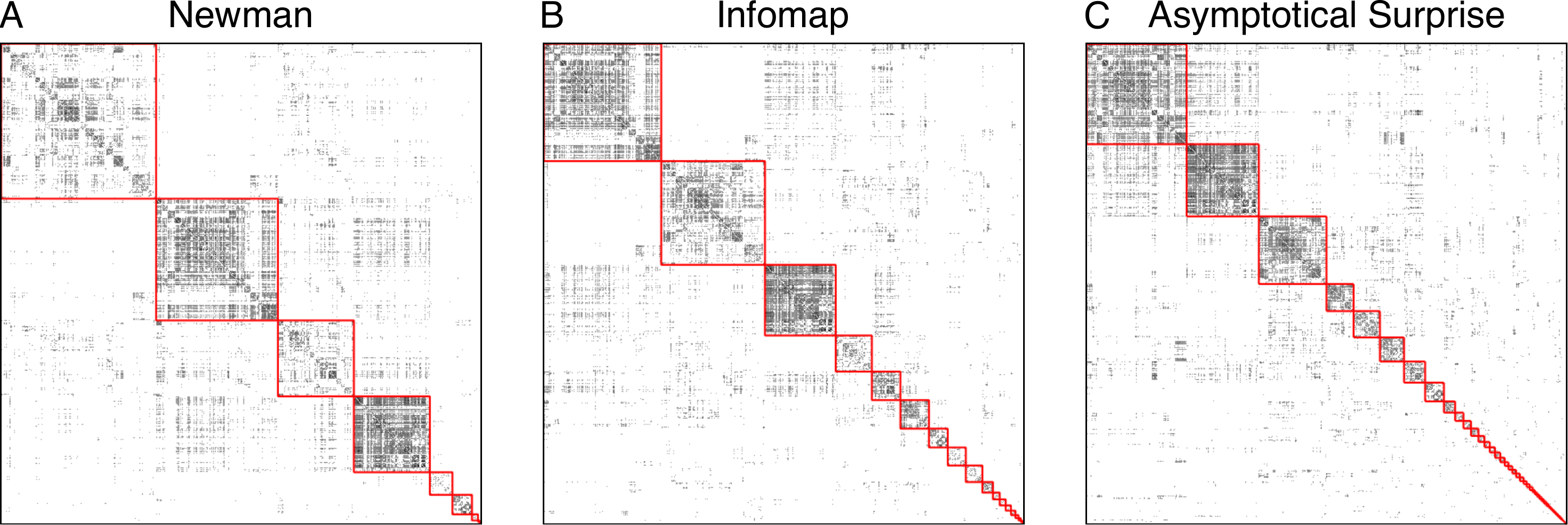}
\caption{A) Louvain-Newman's Modularity partition $Q=0.4967$ B)  Infomap partition $L=8.5173$. C) Asymptotical Surprise partition $S_a=5925.28$.}
\label{fig:partitioncomparison}
\end{figure}

Figure~\ref{fig:partitioncomparison} shows a comparison between the modular structure of the resting state fMRI dataset obtained with Newman's Modularity, Infomap and Asymptotical Surprise.
For each method, we had $10,000$ independent runs and picked the partition with the best value of the respective fitness functions ($Q=0.4967$, $\mathcal{L}=8.5173$, $S_a=5925.3$, for Modularity, Infomap and Asymptotical Surprise, respectively).
The three methods showed significantly different partitions, with a number of detected communities of 10, 19 and 47 for Modularity, Infomap and Asymptotical Surprise, respectively.
Interestingly, Modularity detected a relatively uniform size distribution, consistent with the intrinsic scale built into the fitness function.
Infomap showed a wider distribution of module sizes, with number of nodes ranging between 156 and 3, while Surprise showed the largest spread, and included communities as small as single nodes (singletons).

\begin{figure}[h!]
\centering
\includegraphics[width=\textwidth]{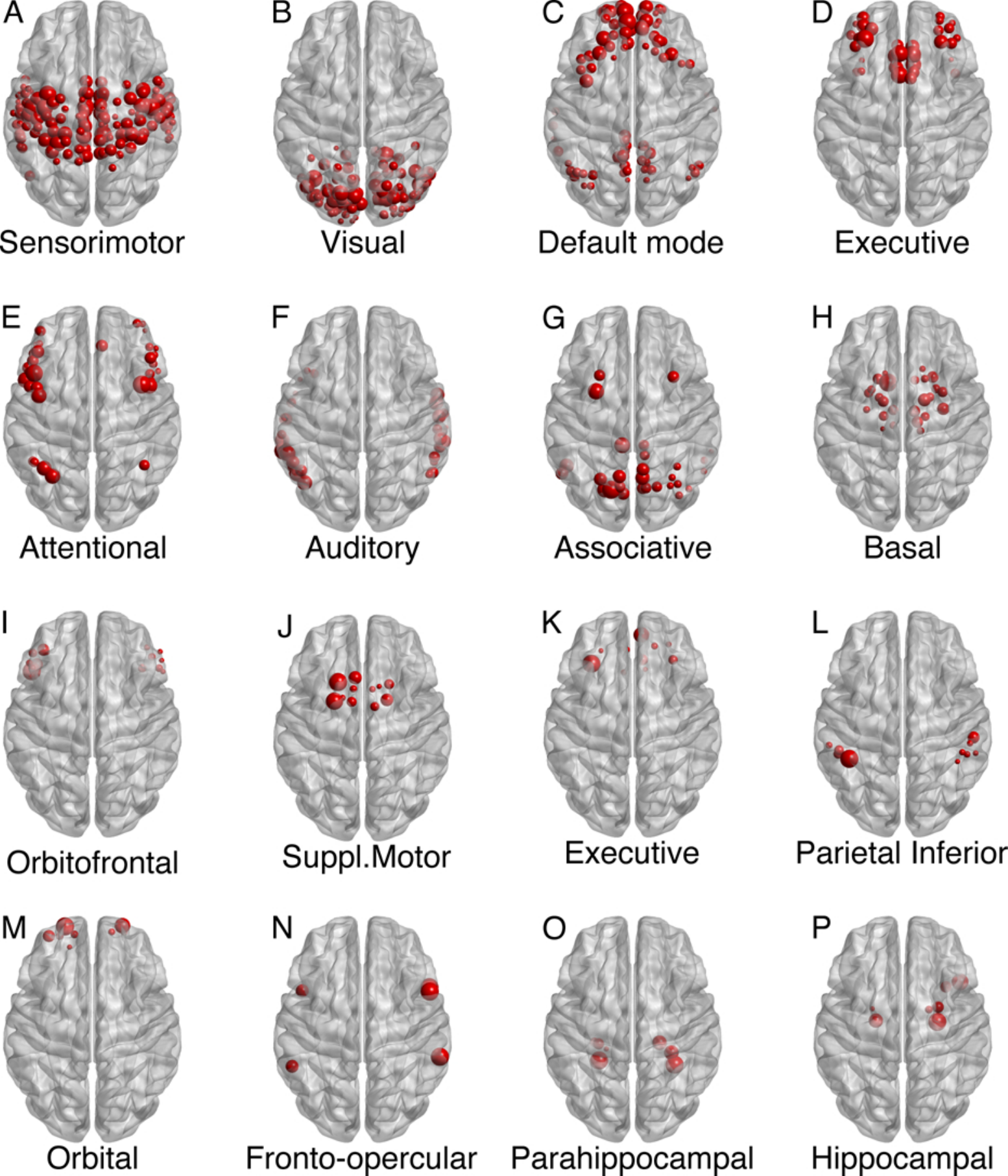}
\caption{Sixteen largest modules found by Asymptotical Surprise Maximization in the resting state network overlaid on an MRI brain template. The module are ranked by decreasing size, and named after corresponding indicative functional networks previously identified by multivariate analysis of resting state fMRI data, or by the comprised anatomical districts.}
\label{fig:cervellini4x4}
\end{figure}

Figure~\ref{fig:cervellini4x4} shows the 16 largest modules detected by Asymptotical Surprise, ranked by number of nodes comprised in each community. 
The first and largest module (Fig. \ref{fig:cervellini4x4}A) includes the pre- and post-central gyri, part of the supramarginal gyrus and supplementary motor area.
The second community (Fig. \ref{fig:cervellini4x4}B) consists largely in nodes belonging to the occipital lobe: the visual areas and the surrounding calcarine sulcus, the lingual and fusiform gyrus.
The third module (Fig. \ref{fig:cervellini4x4}C) reflects the Default Mode Network, spanning the temporo-parietal cortex, the medial prefrontal cortex and the posterior cingulate/precuneus.
The nodes involved in the executive frontal functions form the fourth largest community.
Interestingly, nodes in the communities D,E,G are the major players that take part in the so-called fronto parietal attentional network~\cite{markett2014}.
The auditory network, comprising temporal areas, was detected as a distinct community (Fig. \ref{fig:cervellini4x4}F).
Deeper structures emerge as separate modules in Fig. \ref{fig:cervellini4x4}H, with subcortical areas including the basal ganglia, i.e. putamen, globum pallidum, caudate nucleus and the whole thalamus.
The hippocampus and the parahippocampal gyrus were identified as separate communities (O and P).
Additional, smaller substructures are shown in the third and forth row of Figure~\ref{fig:cervellini4x4}, including the Supplementary Motor Area (Fig. \ref{fig:cervellini4x4}J) and the orbital (Fig. \ref{fig:cervellini4x4}M) and orbitofrontal (Fig. \ref{fig:cervellini4x4}I) modules, containing nodes from Brodmann area 47.

Partitions of the functional connectivity network obtained by Newman's Modularity and Infomap are reported in the Supplementary Materials Section (Figures S4 and S5).
Newman's Modularity retrieved four large, relatively uniform communities, corresponding to the Default Mode Network, the central network, occipital and frontoparietal networks.
This is in keeping with previous studies using  Modularity optimization by spectral decomposition~\cite{crossley2013a}, and consistent with the strong resolution limit that affects this method.
Additionally, a few smaller modules were found by Louvain optimization of Newman's Modularity, corresponding to the basal ganglia, the hippocampal/parahippocampal formation and two asymmetrically distributed subcortical clusters.

Infomap identified 19 communities of various sizes, also shown in the Supplementary Materials Section, Figure S5.
The largest modules showed a close correspondence with those identified by Asymptotical Surprise, albeit with some notable differences.
By way of example, the largest component includes the motor-sensory and auditory modules, identified as separate communities by Asymptotical Surprise.
The Default Mode Network retrieved by Infomap includes parts of the temporal cortices that are not normally associated with the DMN.
Similarly, hippocampus and the parahippocampal modules were merged by Infomap, and resolved as individual modules  by Asymptotical Surprise.
Other modules, including the visual, associative and executive networks (C, E and F in Figure S5, respectively) were qualitatively very similar to those identified by Asymptotical Surprise.

Altogether, the picture that emerges is consistent with the idea that the resolution limit is more severe in Newman's Modularity than in Infomap, and that Asymptotical Surprise presents the best resolving power among the three methods in a real-world network with finite SNR and variability as the resting state functional connectivity network used for this study. 

\section{Discussions}
\subsection{Validation of Asymptotical Surprise in model networks}
The performance of Asymptotical Surprise optimization by PACO was assessed in model graphs with a built-in community structure, and compared with two established community detection methods. We have chosen two synthetic benchmark networks, the ring of cliques and the LFR network.

\begin{figure}[h!]
\centering
\includegraphics[width=\textwidth]{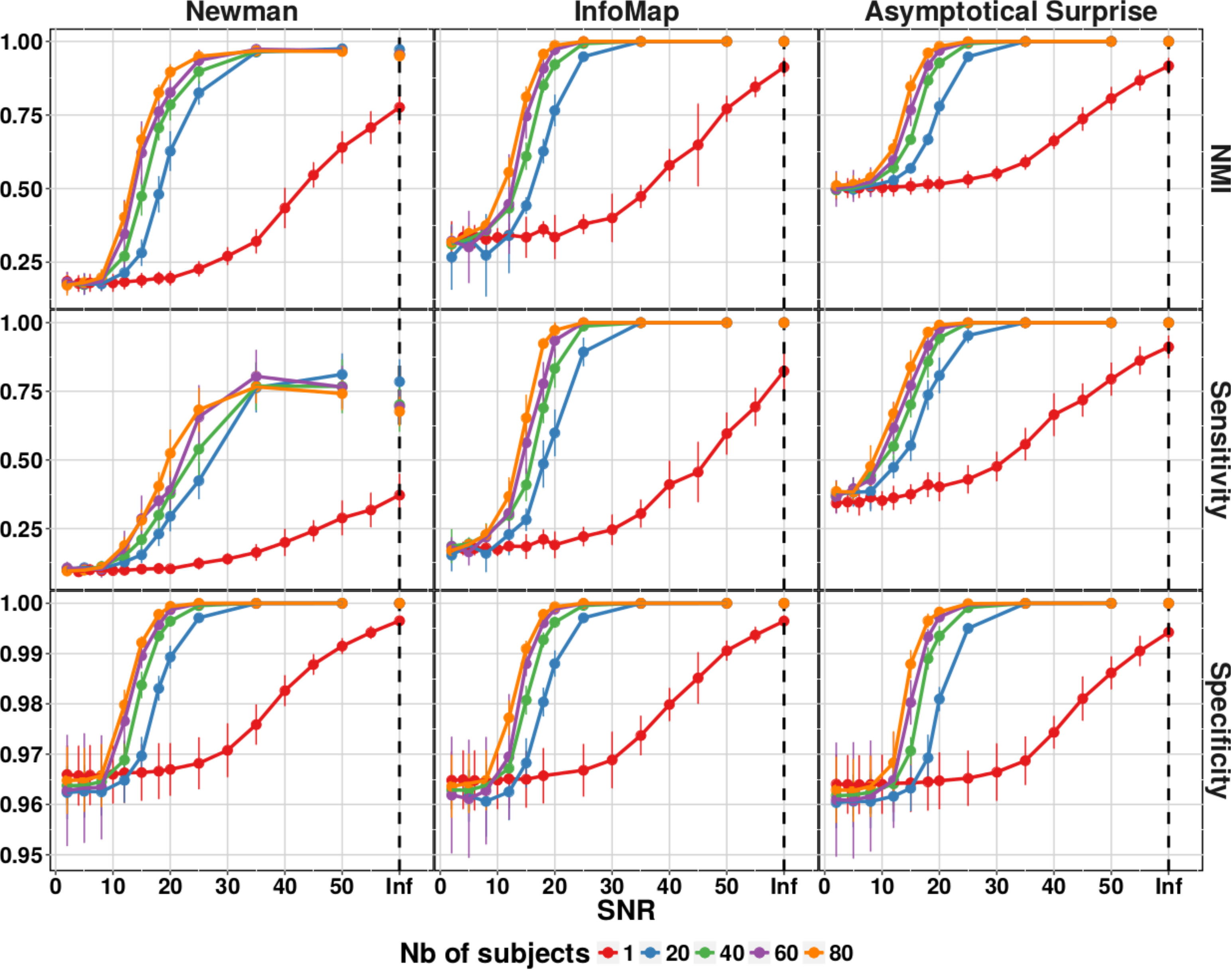}
\caption{NMI, Sensitivity and Specificity of the three community detection algorithms applied to a power-law ring of clique network. SNR indicates Signal to Noise Ratio, and Inf the situation with a network structure unperturbed by noise. Number of Subjects indicates the different number of instances used to generate the group level network.}
\label{fig:nmisensitivityspecificityringclique}
\end{figure}

The ring of cliques presents a clear-cut modular structure by construct, with modules corresponding to complete subgraphs of variable sizes sampled from a power-law distribution. 
This toy network proved useful to assess the effects of the resolution limit in the presence of a wide distribution of cluster sizes. The effects of this limit were particularly apparent for Newman's Modularity (Figure~\ref{fig:nmisensitivityspecificityringclique}), that showed poor Sensitivity even for noiseless rings of cliques, plateauing at a value of $0.75$.
This is consistent with the findings of~\cite{fortunato2007}, that showed that for Modularity the resolution limit is set by the square root of the total number of edges in the graph.
For Infomap, this limit is less severe and is determined by the number of inter-cluster edges~\cite{kawamoto2015}. Accordingly, the effects of the resolution limit were not apparent in this model network, where modules are sparingly connected by single edges.
Asymptotical Surprise presented the best performance, consistent with the idea that this cost function is quasi-resolution limit free~\cite{traag2015}.

However, real brain networks are characterized by heterogeneous distributions of node degree, with fat tails and power-law decays~\cite{bullmore2009}. Such heterogeneity is critical, as it determines some of the remarkable features of brain connectivity networks, including resilience to random failure and rich-clubness~\cite{VandenHeuvel2011,vandenHeuvel2013a}. To provide a more realistic benchmark, we used the Lancichinetti-Fortunato-Radicchi algorithm~\cite{lancichinetti2008}, that makes it possible to generate networks with realistic and tunable power law degree distribution and community sizes.

For LFR networks, the difference in performance in the low-noise regime was more nuanced for the three methods compared in this study, possibly a result of a fuzzier community structure of the LFR network compared to the ring of cliques, and of the narrower distribution of cluster sizes. However, the picture appears different when noise and intersubject variability were injected into the network structure.

Noise and other sources of variability in the data can significantly affect the structure of the resulting network representation.
Noisy fMRI time-courses, for example, may introduce spurious correlations in brain functional connectivity networks.
This problem may be particularly relevant for methods endowed with high resolution, like Asymptotical Surprise, that may be more vulnerable to False Positives generated by the mis-assignment of peripheral nodes, particularly in small clusters. Hence, the resolving power of community detection methods should be gauged against Specificity, which maybe affected by noise in the distribution of edges that define the network's structure.
However, to the best of our knowledge, this aspect has never been considered in the existing literature assessing the performance of community detection algorithms as applied to the study of brain connectivity.

To this end, we have devised methods to inject noise, with amplitude and spectral distribution that mimic those of experimental noise, into networks with a well defined planted structure. Moreover, we have generated difference instances for each network, corresponding to different subjects in a group, to account for intersubject variability that occurs in typical neuroimaging studies. 

Unsurprisingly, for all methods and networks, detection of the planted structure improved with decreasing levels of noise, and with increasing number of subjects in the study.
However, Asymptotical Surprise appeared to provide a superior performance in terms of NMI and Sensitivity to the planted structure for lower SNRs in both types of networks, while its Specificity was in line with that of resolution-limited methods like Newman's and Infomap (Figures \ref{fig:nmisensitivityspecificityringclique},\ref{fig:nmisensitivityspecificitylfr}).
This rules out the idea that the higher sensitivity to small clusters of Asymptotical Surprise may be detrimental in noisy networks, making it more vulnerable to small, spurious modules.

\subsection{Community detection in functional connectivity networks by Asymptotical Surprise}

Application of Asymptotical Surprise maximization to a group-level, resting state functional connectivity network from the brains of $27$ healthy subjects revealed a heterogeneous distribution of modules, with large and small modules coexisting in the optimal partition.
This is in keeping with previous findings with binary Surprise~\cite{Nicolini2016}.
These modules closely reflect functional networks reported in many studies using Independent Component Analysis or other multivariate methods, including the sensorimotor, visual, default mode, executive, and attentional networks.
Moreover, anatomically defined subcortical structures, like the hippocampus and parahippocampal formations emerged as independent moduli.

While this is entirely consistent with our understanding of the neurofunctional and anatomical organization of the human brain, the accuracy of Asymptotical Surprise in identifying these networks is notable.
Indeed, Surprise, like other graph-based community detection methods, divides networks into disjoint clusters of nodes on the basis of topological criteria.
While a correspondence between topological modularity and functional networks identified by, e.g, Independent Component Analysis may be expected, it is not a given, for they are defined on different principles. Indeed, multivariate methods like ICA separate components on the basis of the statistical independence of the time-courses, and do not convey information regarding the mutual relationship between modules nor about their topological organization.

Previous studies applying resolution-limited methods like Newman's Modularity to the same dataset hereby analyzed~\cite{crossley2013a} found a few, large modules encompassing large-scale networks, but failed to identify finer, neurofunctionally  plausible substructures like those shown in the present study. Infomap, on the other hand, proved sensitive to heterogeneously distributed clusters, thus implying that this method does not have an intrinsic scale, like Modularity and variations thereof based on the introduction of a resolution parameter. 
However, Asymptotical Surprise appears to provide superior performance in identifying small subnetworks, particularly in the presence of noise, thus suggesting that this method may represent a new standard for community detection in brain networks. It should also be noted that no symmetry constraint was imposed, and the symmetrical bilateral distribution of nodes in the retrieved modules arises entirely from Asymptotical Surprise optimization. 

Hierarchical clustering methods have been extensively applied to investigate the structure of brain connectivity networks, showing smaller and smaller clusters as the modules are iteratively subdivided~\cite{meunier2010}.  Maximization of Asymptotical Surprise reflects the optimal cut through the dendrogram representing connectivity at these different levels of subdivision, and provides information on the optimal partition of the network. Hence, the heterogeneous distribution of cluster sizes retrieved by Asymptotical Surprise suggests that multiple scales of structure exist at the same level of the dendrogram.

The presence of heterogeneously distributed modules in functional connectivity networks has important consequences for our understanding of the brain functional organization. By way of example, it has been shown that highly connected nodes, or hubs, are critically important in brain connectivity networks, and may play different roles depending on their position and connectivity distribution within and between modules~\cite{bullmore2009}. Hubs that primarily connect to nodes within the same community are dubbed “provincial hubs”, and are thought to be responsible for the definition and stability of the modules. Conversely, hubs that connect different modules are referred to as “connector hubs” and ensure integration of the activity of the network. The classification of hubs  strongly depends on the modular structure that is considered, and inaccurate partitioning due to the resolution limit can lead to the wrong interpretation of their role in the interplay between segregation and integration of brain function~\cite{bullmore2009}. The present study suggests that this may have been the case in previous studies, in which resolution limited methods characterized by an intrinsic scale have been used, and provides a solution that will enable more accurate classification of hubs and nodes.

Finally, abnormal functional connectivity has been observed in a number of neurological and psychiatric diseases, but the coarse resolution of methods like Newman's Modularity~\cite{Fornito2015} may have not detected differences in the modular organization of networks in patients compared to healthy controls. The improved resolution and sensitivity to multiscale structure afforded by Asymptotical Surprise may provide a powerful means to assess the brain functional architecture in disease states, thus contributing a potential imaging-based marker and a key to interpret the functional effects of aberrant connectivity.

\subsection{Limitations}
Some caution should be taken in the interpretation of the graphs in Figures \ref{fig:nmisensitivityspecificityringclique} and \ref{fig:nmisensitivityspecificitylfr}.
Indeed, the SNRs of the synthetic networks we have generated reflect noise with features, like a Rician distribution, that mimic some, but not all aspects of the variability of experimental data.
By way of example, the brain parcellation scheme applied to define the nodes, and the heterogeneity of voxels within these parcels, may play a role that is difficult to model in toy networks~\cite{Fornito2010}.
Hence, the simulated Sensitivity and Specificity as a function of SNR and number of subjects should not be taken as absolute values to be used in the power and sample size estimation in real experimental designs.
Nevertheless, these simulations provide useful information on the dependence of these parameters on noise levels, and a rigorous means to assess the relative merits of different community detection methods.

 Finally, we should note that the maximum value of Asymptotical Surprise calculated with PACO is an index of quality of the entire partition, and not of individual modules.
Hence, individual modules may not all have the same strength of internal cohesiveness relative to their connection with other modules. We have found hints of this phenomenon in the comparison of nearly-optimal partitions obtained in the $10,000$ runs of PACO that we have performed to find the optimal community structure for this network.
The overall community structure appeared to be robust, with most modules persistently emerging in every nearly-optimal partition, but in some cases we observed pairs of modules splitting or merging in otherwise similar solutions. Most notably, this was observed for the thalamus that in some instances was merged with the basal cluster and in others featured as a separate module.
This phenomenon may be less critical for methods like Newman's Modularity that have an intrinsic scale and retrieve uniformly distributed modules.

\section{Conclusion}
We have extended the use of Surprise, a resolution-limit-free fitness function for the study of the modular structure of complex networks, to weighted brain functional connectivity networks. Specifically, we have developed a novel method, dubbed PACO, for the optimization of Asymptotical Surprise, a weighted counterpart of Surprise in the limit of large networks. We have applied PACO optimization of Asymptotical Surprise in synthetic networks to evaluate the relative merits of this novel approach against Newman's Modularity and Infomap, two of the leading methods used for community detection in brain connectivity networks. Specifically, we have implemented a process to inject noise into networks endowed with a ground-truth modular structure to assess the trade-off between improved resolution afforded by Asymptotical Surprise and potential sensitivity to spurious correlations introduced by variability in the data. Asymptotical Surprise optimization proved superior to existing methods in terms of Sensitivity and accuracy in detection of the planted structure as measured by Normalized Mutual Information, while showing comparable Specificity. We have also applied our approach to the partitioning of functional connectivity networks from resting state fMRI experiments. Direct comparison with other methods clearly demonstrated improved capability to identify neurofunctionally plausible and anatomically well-defined substructures otherwise concealed by the resolution limit. Asymptotical Surprise revealed a complex modular structure of resting state connectivity, with communities of widely different sizes reflecting distributed functional networks alongside with small, anatomically or functionally defined modules. This evidence corroborates the idea that the resolution limit has negatively affected current models of the brain modular organization and the identification of the hubs responsible for integration of functional modules.  The application of methods like Asymptotical Surprise provides a novel, powerful approach to study the modular structure of brain connectivity beyond this limit.

\section*{Acknowledgments}
The authors wish to thank Prof. Edward Bullmore and Prof. Nicholas Crossley for providing data and templates, and Prof. Pasquina Marzola for her continuing support. This project has received funding from the European Union's Horizon 2020 research and innovation program under grant agreement No 668863.

\section*{Author contributions}
C.B, C.N, A.B. conceived the experiments, C.N., C.B. conducted the experiments, C.N. C.B, and A.B. wrote the paper and reviewed the analysis. All authors reviewed the manuscript.
Our implementation of PACO as well as the code to generate benchmark LFR networks was written in \texttt{C++} with bindings in Matlab, Octave,  Python. This software is available upon request.

\subsection*{Competing financial interests}
The authors declare no competing financial interests.

\newpage
\section*{Figures}

\newpage

\newpage

\newpage

\newpage


\newpage
\begin{table}
\centering
\begin{tabular}{|c|c|c|c|}
\hline
\textbf{NMI} & Newman & Infomap & Asymptotical Surprise \\ \hline 
Newman & 1.00 & 0.75 & 0.62 \\ \hline 
Infomap & - & 1.00 & 0.76 \\ \hline 
Asymptotical Surprise & - & - & 1.00 \\ \hline 
\end{tabular}
\caption{NMI values of partitions obtained from the three different methods on the resting state network.}
\label{tab:nmimethods}
\end{table}

\end{document}